\documentclass[fleqn,a4paper]{article}
\usepackage{a4wide}
\usepackage{amsmath}
\usepackage{amssymb}
\usepackage{comment}
\usepackage{hyphenat}
\usepackage{multicol}
\usepackage{times}
\usepackage{bm}
\usepackage{dsfont}
\usepackage{caption}
\usepackage{graphicx}
\usepackage{hyperref}

\usepackage{natbib}
\def\T{{ \mathrm{\scriptscriptstyle T} }}

\usepackage[plain,vlined]{algorithm2e}

\makeatletter
\renewcommand{\algocf@captiontext}[2]{#1\algocf@typo. \AlCapFnt{}#2} 
\def\@algocf@capt@plain{top}
\renewcommand{\algocf@makecaption}[2]{%
	\addtolength{\hsize}{\algomargin}%
	\sbox\@tempboxa{\algocf@captiontext{#1}{#2}}%
	\ifdim\wd\@tempboxa >\hsize
	\hskip .5\algomargin%
	\parbox[t]{\hsize}{\algocf@captiontext{#1}{#2}}
	\else%
	\global\@minipagefalse%
	\hbox to\hsize{\box\@tempboxa}
	\fi%
	\addtolength{\hsize}{-\algomargin}%
}
\makeatother

\newtheorem{theorem}{Theorem}
\newtheorem{definition}{Definition}
\newtheorem{proposition}{Proposition}
\newtheorem{lemma}{Lemma}

\newcommand{\qed}{\hfill$\Box$}
\newenvironment{proof}[1][]{\addvspace{.2cm} \noindent{\bf Proof #1}}{\qed\vspace{.3cm}}

\title{Bayesian cumulative shrinkage for infinite factorizations}
\author{Sirio Legramanti\footnote{Department of Decision Sciences, Bocconi University, 20136 Milan, Italy, \texttt{sirio.legramanti@unibocconi.it, daniele.durante@unibocconi.it}} \and Daniele Durante\footnotemark[\value{footnote}] \and David B. Dunson\footnote{Department of Statistical Science, Duke University, Durham, NC 27708, U.S.A., \texttt{dunson@duke.edu}}}

\begin{document}
\maketitle

\begin{abstract}
There is a wide variety of models in which the dimension of the parameter space is unknown.  For example, in factor analysis the number of latent factors is typically not known and has to be inferred from the observed data. Although classical shrinkage priors are useful in these contexts, increasing shrinkage priors can provide a more effective option, which progressively penalizes expansions with growing complexity.  In this article we propose a novel increasing shrinkage prior, named the cumulative shrinkage process, for the parameters controlling the dimension in over-complete formulations. Our construction has broad applicability, simple interpretation, and is based on a sequence of spike and slab distributions which assign increasing mass to the spike as model complexity grows.   Using factor analysis as an illustrative example, we show that this formulation has theoretical and practical advantages over current competitors, including an improved ability to recover the model dimension.  An adaptive Markov chain Monte Carlo algorithm is proposed, and the methods are evaluated in simulation studies and applied to personality traits data. Code is
available at \url{https://github.com/siriolegramanti/CUSP}.\\

\noindent	\textit{Some key words:} Factor analysis; 
Increasing shrinkage;
Multiplicative gamma process;  
Spike and slab;
Stick-breaking
\end{abstract}

\section{Introduction}
\label{sec_intro}
There has been a considerable interest in shrinkage priors for high dimensional parameters \citep[e.g.,][]{ishwaran2005spike,carvalho2010} but most of the focus has been on regression, where there is no natural ordering in the coefficients. There are several settings, however, where an order is present and desirable.  Indeed, in statistical models relying on low-rank factorizations or basis expansions, such as factor models and tensor factorizations, it is natural to expect that additional dimensions play a progressively less important role in characterizing the data or model structure, and hence the associated parameters should have a stochastically decreasing effect. Such a behavior can be induced through increasing shrinkage priors. For instance, in the context of Bayesian factor models an example of this approach can be found in the multiplicative gamma process developed by \citet{bhattacharya2011sparse} to penalize the effect of additional factor loadings via a cumulative product of gamma priors for their precision. Although this prior has been widely applied, there are practical disadvantages that motivate consideration of alternative solutions \citep{durante2017note}.  In general, despite the importance of increasing shrinkage priors in many factorization models, the methods, theory and computational strategies for these priors remain under-developed.


Motivated by the above considerations, we propose a novel increasing shrinkage prior, the cumulative shrinkage process, which is broadly applicable, while having simple and parsimonious structure. The proposed prior induces increasing shrinkage via a sequence of spike and slab distributions assigning growing mass to the spike as model complexity grows. In Definition~\ref{def_cusp}, we present this prior for the general case in which the effect of the $h$th  dimension is controlled by a scalar parameter $\theta_h \in \Re$, so that redundant terms can be essentially deleted by progressively shrinking the sequence $\theta=\{ \theta_h \in \Theta \subseteq \Re: h=1, 2, \ldots\}$ towards an appropriate value $ \theta_\infty \in \Re$. For example, in factor models $\theta_h\in \Re_+$ may denote the variance of the loadings for the $h$th factor, and the goal is to define a prior on these terms which favors stochastically decreasing impact of the factors via increasing concentration of the loadings near zero as $h$ grows.

\begin{definition} \label{def_cusp}
	Let $\theta=\{ \theta_h \in \Theta \subseteq \Re: h=1, 2, \ldots\}$ denote a countable sequence of parameters. We say that $\theta$ is distributed according to a cumulative shrinkage process with parameter $ \alpha>0 $, starting slab distribution $P_0$ and target value $\theta_\infty$ if, conditionally on $ \pi = \{ \pi_h \in (0,1): h=1, 2, \ldots\} $, each $ \theta_h $ is independent and has the following spike and slab distribution:
	\begin{eqnarray} 
	(\theta_{h} \mid \pi_h) \sim P_h = ( 1 - \pi_h )  P_0 + \pi_h  \delta_{\theta_\infty}, 
	\qquad \pi_h = \sum\nolimits_{l=1}^{h} \omega_l, \qquad \omega_l= v_l \prod\nolimits_{m=1}^{l-1} (1-v_m), 
	\label{eq_cusp}
	\end{eqnarray}
	where $v_1, v_2, \ldots$ are independent  $\mbox{\normalfont Beta}(1,\alpha)$ variables and $P_0$ is a diffuse continuous distribution.
\end{definition}

Equation~\eqref{eq_cusp} exploits the stick-breaking construction of the Dirichlet process \citep{ishwaran_2001}. This implies that the probability $ \pi_h $ assigned to the spike $\delta_{\theta_\infty}$ increases with the model dimension $h$, and that $\lim_{h \to \infty} \pi_h = 1$ almost surely.  Hence, as complexity grows, $P_h$ increasingly concentrates around $\theta_\infty$,  which is specified to facilitate the deletion of redundant terms, while the slab $P_0$ corresponds to the prior on the active parameters. Definition~\ref{def_cusp} can be extended to sequences in $ \Re^p $, and $\delta_{\theta_\infty}$ can be replaced with a continuous distribution, without affecting the key properties of the prior, which are presented in \S~\ref{sec_cusp}. As we will discuss in \S~\ref{sec_cusp} and in \S~\ref{subsec_factor_formulation}, it is also possible to restrict  Definition~\ref{def_cusp} to finitely many terms $(\theta_1, \ldots, \theta_H)$ by letting $v_H=1$. In practical implementations, this truncated version typically ensures full flexibility if $H$ is set to a conservative upper bound, but this value can be extremely large in several  high dimensional settings,  thus motivating our initial focus on the  infinite expansion and its theoretical properties.

\section{General properties of the cumulative shrinkage process}
\label{sec_cusp}
We first motivate our cumulative stick-breaking construction for the sequence $\pi$ that controls the mass assigned to the spike in \eqref{eq_cusp} as a function of  model dimension. Indeed, one could alternatively consider pre-specified non-decreasing functions bounded between $0$ and $1$.  However, we have found that such specifications are overly-restrictive and have  worse practical performance.  The specification in  \eqref{eq_cusp} is purposely chosen to be effectively nonparametric, with Proposition~\ref{prop_step} showing that the prior has large support on the space of non-decreasing sequences taking values in $(0,1)$.
See the Appendix for proofs.


\begin{proposition} \label{prop_step}
	Let $ \Pi $ be the probability measure induced on  $\pi = \{ \pi_h \in (0,1): h=1, 2, \ldots\} $ by~\eqref{eq_cusp}, then $ \Pi $ has large support on the whole space of non-decreasing sequences taking values in $ (0,1)$.
\end{proposition}
Besides being fully flexible, our   construction for $\pi$  also has simple interpretation and allows control over shrinkage via an interpretable parameter $\alpha$, as stated in Proposition~\ref{lemma_dtv} and in the subsequent results.

\begin{proposition} \label{lemma_dtv}
	Each $ \pi_h $ in~\eqref{eq_cusp} coincides with the proportion of the total variation distance between the slab and the spike covered up to step $ h $, in the sense that $\pi_h= d_{\textsc{tv}}(P_0,P_h) / d_{\textsc{tv}}(P_0, \delta_{\theta_\infty})$ for every $h$.
\end{proposition}

Using similar arguments, we can obtain analogous expressions for $ \omega_h $ and $ v_h $, which represent the proportions of the total $d_{\textsc{tv}}(P_0, \delta_{\theta_\infty})$ and the remaining $d_{\textsc{tv}}(P_{h-1}, \delta_{\theta_\infty})$, respectively, covered between steps $h-1$ and $h$. Specifically, $\omega_h=d_{\textsc{tv}}(P_{h-1},P_h)/d_{\textsc{tv}}(P_0, \delta_{\theta_\infty})$ and $v_h=d_{\textsc{tv}}(P_{h-1}{,}P_h)/d_{\textsc{tv}}(P_{h-1}, \delta_{\theta_\infty})$ for every $h$. The expectations of these quantities are explicitly available as
\begin{eqnarray} 
E(v_h)=\frac{1}{1+\alpha}, 
\qquad  E(\omega_h)=\frac{\alpha^{h-1}}{(1+\alpha)^h}, \qquad  
E (\pi_h) =  1 - \frac{\alpha^h}{(1+\alpha)^h} 
\qquad (h=1, 2, \ldots).
\label{eq_cusp_mean}
\end{eqnarray}
Moreover, combining~\eqref{eq_cusp_mean} with Definition~\ref{def_cusp}, the expectation of $ \theta_h \ (h=1,2,\ldots)$ is
\begin{eqnarray} 
E(\theta_h) = E\{E(\theta_h \mid \pi_h)\}=\{1-E(\pi_h ) \}  \theta_{0}+E( \pi_h)  \theta_{\infty}=\theta_{\infty} + \{\alpha(1+\alpha)^{-1}\}^h(\theta_0- \theta_{\infty}), 
\label{eq_cusp_theta}
\end{eqnarray}
where $\theta_{0}$ defines the expected value under the slab  $P_0$.  Hence, as $h$ grows, the prior expectation of $\theta_h$  converges exponentially towards the spike location $\theta_{\infty}$.  As stated in Lemma \ref{lemma_ordering}, a stronger notion of cumulative shrinkage in distribution, beyond simple concentration in expectation, also holds under \eqref{eq_cusp}.

\begin{lemma}\label{lemma_ordering}
	Let ${\mathbb{B}}_{\varepsilon}(\theta_{\infty})=\{\theta_h \in \Theta \subseteq \Re: |\theta_h-\theta_{\infty}| \leq \varepsilon \}$ denote an  $\varepsilon$-neighborhood around $\theta_{\infty}$ with radius $ \varepsilon > 0$, and define with $\bar{\mathbb{B}}_{\varepsilon}(\theta_{\infty})$ the complement of ${\mathbb{B}}_{\varepsilon}(\theta_{\infty})$. Then, for any $h=1, 2, \ldots$ and $\varepsilon > 0$,
	\begin{equation}
	\mbox{\normalfont pr}(|\theta_h-\theta_{\infty}| > \varepsilon) =
	P_0\{\bar{\mathbb{B}}_{\varepsilon}(\theta_{\infty})\} \{\alpha(1+\alpha)^{-1}\}^h.
	\label{eq_prob_ball}
	\end{equation}
	Therefore, $\mbox{\normalfont pr}(|\theta_{h+1}{-} \ \theta_{\infty}| \leq \varepsilon) > \mbox{\normalfont pr}(|\theta_h{-} \ \theta_{\infty}| \leq \varepsilon)$ for any $\alpha>0$, $h=1, 2, \ldots$ and $\varepsilon > 0$.
\end{lemma}

Equations~\eqref{eq_cusp_mean}--\eqref{eq_prob_ball} highlight how the rate of increasing shrinkage is controlled by $\alpha$.  In particular, lower values of $\alpha$ induce faster concentration around $\theta_\infty$ and hence more rapid shrinkage of the redundant terms. This control over the rate of increasing shrinkage via $\alpha$ is separated from the specification of the slab~$P_0$, thereby allowing flexible modelling of the active terms. As discussed in \citet{durante2017note}, such a separation does not hold, for example, in the multiplicative gamma process  \citep{bhattacharya2011sparse} whose hyper-parameters control both the rate of shrinkage and the prior for the active factors. This creates a trade-off between the need to maintain diffuse priors for the active terms and the attempt to shrink the redundant ones.   Moreover, increasing shrinkage holds only in expectation and for specific hyper-parameters.

Instead,  our prior ensures increasing shrinkage in distribution for any $\alpha$, and can model any prior expectation on the number of active terms. In fact, $\alpha$ is equal to the prior mean of the number of terms in $\theta$ modelled via the slab $P_0$. This result follows after noticing that $ (\theta_h \mid \pi_h)$ in~\eqref{eq_cusp} can be alternatively obtained by marginalizing out the augmented indicator $ c_h \sim \mbox{Bern}(1-\pi_h)$ in  $(\theta_h \mid c_h) \sim c_h P_0 +(1-c_h)\delta_{\theta_\infty}$. According to this result, $H^*=\sum\nolimits_{h=1}^\infty c_h$ counts the number of active elements in $ \theta $, and its prior mean is
\begin{eqnarray*}
	E(H^*)
	\ {=}\sum\nolimits_{h=1}^\infty E(c_h)
	\ {=}\sum\nolimits_{h=1}^\infty E\{E(c_h \mid \pi_h)\}
	\ {=}\sum\nolimits_{h=1}^\infty E(1-\pi_h)
	\ {=}\sum\nolimits_{h=1}^\infty\{\alpha(1+\alpha)^{-1}\}^h
	\ {=} \ \alpha.
\end{eqnarray*}
Hence, $ \alpha $ should be set to the expected  number of active terms, while $P_0$ should be sufficiently diffuse to model active components, and $\theta_{\infty}$ should be chosen to facilitate the deletion of redundant ones.

Recalling \cite{bhattacharya2011sparse} and \cite{rousseau2011}, it is useful to define models with more than enough components and then choose shrinkage priors which favor effective deletion of the unnecessary ones. This choice protects against over-fitting and allows estimation of model dimension, bypassing the need for reversible jump \citep{lopes2004} or other computationally intensive strategies.  Our cumulative shrinkage process in \eqref{eq_cusp} provides a useful prior for this purpose.  As discussed in \S~\ref{sec_intro}, it is straightforward to modify Definition \ref{def_cusp} to instead restrict to $H$ components, by letting $v_H=1$, with $H$ a conservative upper bound.  Theorem~\ref{thm_trunc2} provides theoretical support for such a truncated representation.


\begin{theorem} \label{thm_trunc2}
	If $\theta$ has prior~\eqref{eq_cusp} and $\theta^{(H)}$ denotes the sequence obtained by fixing $ \theta_h=0 $ in $ \theta $ for every $ h>H $, then for any truncation index $ H $ and $\varepsilon \geq |\theta_{\infty}|$,
	\begin{equation*} \label{eq_trunc3}
	\mbox{\normalfont pr}\{d_{\infty}(\theta,\theta^{(H)})>\varepsilon\}=\mbox{\normalfont pr}\{\sup(|\theta_{h}|: h=H+1, H+2, \ldots)>\varepsilon\} \leq P_0\{\bar{\mathbb{B}}_{\varepsilon}(0)\}   \alpha\{\alpha(1+\alpha)^{-1}\}^H,
	\end{equation*}
	where $ d_\infty $ is the sup-norm distance and $\bar{\mathbb{B}}_{\varepsilon}(0)$ is the complement of ${\mathbb{B}}_{\varepsilon}(0)=\{\theta_h \in \Theta \subseteq \Re: |\theta_h| \leq \varepsilon \}$.
\end{theorem}
Hence,  the prior probability of $\theta^{(H)}$ being close to $\theta$ converges to one at a rate which is exponential in $H$, thus justifying posterior inference under finite sequences based on a conservative $H$. Although the above bound holds for $\varepsilon \geq |\theta_{\infty}|$, in general $\theta_{\infty}$ is set close to zero. Hence, Theorem~\ref{thm_trunc2} is valid also for small $\varepsilon$.

\section{Cumulative shrinkage process for  Gaussian factor models}
\label{sec_factor}
\subsection{Model formulation and prior specification}
\label{subsec_factor_formulation}

Definition~\ref{def_cusp} provides a general prior which can be used in different models \citep[e.g.,][]{gopalan2014bayesian} under appropriate choices of $P_0$  and $\theta_{\infty}$. Here, we focus on Gaussian sparse factor models as an important special case to illustrate our approach.  We will compare primarily to the multiplicative gamma process, which has been devised specifically for this class of models  and was shown to have practical gains  in this context relative to several competitors, including the use of lasso \citep{tibshirani1996}, elastic-net \citep{zou2005} and banding approaches \citep{bickel_2008}.  Although there are other priors for sparse factor models \citep[e.g.,][]{carvalho_2008, knowles_2011}, these choices have practical disadvantages relative to the multiplicative gamma process, so they will not be considered further here.


The focus will be on performance in learning the structure of the $p \times p$ covariance matrix $\Omega=\Lambda \Lambda^{\T}+ \Sigma$ for the data  $ y_i=(y_{i1}, \ldots, y_{ip})^{\T} \in \Re^p$ generated from the Gaussian factor model $ {y_{i}=\Lambda \eta_i+\epsilon_i} $, with $\eta_{ih} \sim N(0,1)$, $(i=1, \ldots, n; h=1, 2, \ldots)$, $\epsilon_i \sim N_p(0, \Sigma)$ $(i=1, \ldots, n)$ and $ {\Sigma = \mbox{diag}(\sigma_1^2,\ldots,\sigma_p^2)}$. To perform Bayesian inference for this model, \citet{bhattacharya2011sparse} assumed  $\sigma^2_{j} \sim \mbox{InvGa}(a_{\sigma},b_{\sigma})$ $(j=1, \ldots, p)$, and   $(\lambda_{jh} \mid \phi_{jh},\theta_{h}) \sim N(0,\phi_{jh}\theta_{h})$ $(j=1, \ldots, p;  h=1, 2, \ldots)$ with  scales $\phi_{jh}$ from independent $\mbox{InvGa}(\nu/2,\nu/2)$ priors and global precisions $\theta_h^{-1}$ having multiplicative gamma process prior
\begin{eqnarray} \label{eq_mgp}
\theta_h^{-1}=\prod\nolimits_{l=1}^h \vartheta_l \quad (h=1, 2, \ldots), \qquad \vartheta_1 \sim \mbox{Ga}(a_1, 1), \qquad \vartheta_l \sim \mbox{Ga}(a_2, 1) \ \ (l=2, 3, \ldots).
\end{eqnarray}
Specific choices of $(a_1,a_2)$ in \eqref{eq_mgp} ensure that $E(\theta_h)$ decreases with $h$, thus allowing increasing shrinkage of the loadings as $h$ grows. Instead, we keep $\sigma^2_{j} \sim \mbox{InvGa}(a_{\sigma},b_{\sigma})$ $(j=1, \ldots, p)$, but let $(\lambda_{jh} \mid \theta_h) \sim N(0,\theta_h) $ $ (j=1, \ldots, p; \ h=1, 2, \ldots)$ and place our cumulative shrinkage process prior on $ \theta_h $ by assuming
\begin{eqnarray} \label{eq_cusp_fact}
(\theta_{h} \mid \pi_h) \sim ( 1 - \pi_h ) \mbox{InvGa}(a_\theta,b_\theta) + \pi_h  \delta_{\theta_\infty}, 
\quad \pi_h = \sum\nolimits_{l=1}^{h} \omega_l, \quad \omega_l= v_l \prod\nolimits_{m=1}^{l-1} (1-v_m), 
\end{eqnarray}
where $v_1, v_2, \ldots$ are independent  $\mbox{\normalfont Beta}(1,\alpha)$.  Integrating out $\theta_h$,  each loading $\lambda_{jh}$ has the marginal prior $( 1 - \pi_h ) t_{2a_\theta}(0,b_\theta / a_\theta) + \pi_h  N(0,\theta_\infty)$, where $ t_{2a_\theta}(0,b_\theta / a_\theta)$ denotes the Student-$t$ distribution with $2a_\theta$ degrees of freedom, location $0$ and scale $b_\theta / a_\theta$. Hence,  $ \theta_\infty $ should be set close to zero to allow effective shrinkage of redundant factors, while $(a_\theta,b_\theta)$ should be specified so as to induce a moderately diffuse prior with scale $b_\theta / a_\theta$ for the active loadings. Although the choice $\theta_{\infty}=0$ is possible, we follow \citet{ishwaran2005spike} by suggesting $\theta_{\infty}>0$ to induce a continuous shrinkage prior on every $\lambda_{jh}$ which improves mixing and  identification of the inactive factors. Exploiting the marginals for $\lambda_{jh}$, it also follows that, if $b_\theta / a_\theta > \theta_\infty$ then $\mbox{\normalfont pr}(|\lambda_{j,h+1}| \leq \varepsilon) > \mbox{\normalfont pr}(|\lambda_{jh}| \leq \varepsilon)$ for each $j=1, \ldots, p$, $h=1, 2, \ldots$  and $\varepsilon >0$. This allows cumulative shrinkage in distribution  also for the loadings, and provides guidelines on $(a_\theta, b_\theta)$ and $\theta_{\infty}$. Additional discussion on prior elicitation and empirical studies on sensitivity can be found in \S~\ref{subsec_factor_empirical}.

To implement the analysis, we require a truncation $H$ on the number of factors needed to characterize $\Omega$, as discussed in \S~2.  Theorem~\ref{thm_support} states that our shrinkage process truncated at $H$ terms induces a well-defined prior for $\Omega$ with full-support, under the sufficient conditions that $ H $ is greater than the {\em true} $ H_0 $, and $E(\theta_h)<\infty$. These conditions are met when considering up to $p$ active factors, with $a_{\theta}>1$ and $\theta_{\infty}<\infty$.


\begin{theorem} \label{thm_support}
	Let $ \Omega_0 $ be any $ p \times p $ covariance matrix and define with $ \Pi $ the prior probability measure on $ p \times p $ covariance matrices $\Omega$ induced by a Bayesian factor model having prior~\eqref{eq_cusp_fact} on $\theta$, truncated at $ H $ with $ v_H = 1 $. If $E(\theta_h)<\infty$, then $ {\Pi \{ \Omega \in \Re^{p \times p}: \Omega\mbox{ has finite entries and is positive semi-definite} \} = 1 }$. In addition, if there exists a decomposition $\Omega_0=\Lambda_0 \Lambda_0^{\T} + \Sigma_0$, such that $ \Lambda_0 \in \Re^{p \times H_0}$ and $ H_0 < H $, then $ \Pi\{ B_\varepsilon^\infty(\Omega_0) \} >0 $ for any $ \varepsilon>0 $, where $ B_\varepsilon^\infty(\Omega_0) $ is an $ \varepsilon $-neighborhood of $ \Omega_0 $ under the sup-norm.
\end{theorem}

Recalling  Theorem 2 in \citet{bhattacharya2011sparse}, this result is also sufficient to ensure that the posterior  of $\Omega$ is weakly consistent  \citep{schwartz1965bayes}. 

\subsection{Posterior computation via Gibbs sampling}\label{subsec_factor_gibbs}
Posterior inference for the factor model in \S~\ref{subsec_factor_formulation} with cumulative shrinkage process \eqref{eq_cusp_fact} truncated at $H$ terms for the loadings, proceeds via a Gibbs sampler cycling across the steps in Algorithm~\ref{alg_factor}.  This sampler relies on a data augmentation which exploits the fact that prior  \eqref{eq_cusp_fact}  can be obtained by marginalizing out the independent indicators  $z_h \ (h=1,\ldots,H)$ with probabilities $\mbox{pr}(z_h=l \mid \omega_l)=\omega_l$ $(l=1, \ldots, H)$ in
\begin{eqnarray}
\label{eq_factor_z}
(\theta_{h} \mid z_h) \sim \{ 1- \mathds{1}(z_h \leq h)\}   \mbox{InvGa}(a_\theta,b_\theta) + \mathds{1}(z_h \leq h)  \delta_{\theta_\infty},
\end{eqnarray}
where $\mathds{1}(z_h \leq h)=1$ if $z_h \leq h$ and $0$ otherwise. As is clear from Algorithm~\ref{alg_factor}, conditioned on $z_1, \ldots, z_H$, it is possible to sample from conjugate full-conditionals, whereas the updating of the augmented data relies on the full-conditional distribution
\begin{eqnarray}
\mbox{pr}(z_h=l \mid -)\propto \left\{ \begin{array}{ll}
\omega_l N_p(\lambda_h; 0, \theta_\infty I_p ), &  \qquad\text{for \ $l =1,\dots,h,$}\\
\omega_l  t_{2a_\theta}\{\lambda_h; 0, (b_{\theta}/a_{\theta}) I_p \}, & \qquad \text{for \ $l = h+1, \dots, H,$}\\
\end{array} \right. 
\label{eq_discr_1}
\end{eqnarray}
where $ N_p(\lambda_h; 0, \theta_\infty I_p )$ and $t_{2a_\theta}\{\lambda_h; 0, (b_{\theta}/a_{\theta}) I_p \} $ are the densities of $p$-variate Gaussian and Student-$t$ distributions, respectively, evaluated at $\lambda_h=(\lambda_{1h}, \ldots, \lambda_{ph})^{\T}$. Equations~\eqref{eq_discr_1} are obtained by marginalizing out $ \theta_h $, distributed as in~\eqref{eq_factor_z}, from the joint $ N_p(\lambda_h; 0, \theta_h I_p) $. These calculations are straightforward in a variety of Bayesian models based on conditionally conjugate constructions, thus making \eqref{eq_cusp} a general prior which can be easily incorporated, for instance, in Poisson factorizations \citep{gopalan2014bayesian}. 

\begin{algorithm}[t]
	\caption{One cycle of the Gibbs sampler for factor models with the cumulative shrinkage process} 
	\label{alg_factor}
	\vspace{-7pt}
	\nl \For{j from 1 to p}{sample the $j$th row of $\Lambda$ from $ N_H (V_j \eta^{\T}\sigma^{-2}_jy_j, V_j ) $, with $V_j=(D^{-1}+\sigma^{-2}_j\eta^{\T}\eta)^{-1}$, $D=\mbox{diag}(\theta_1, \ldots, \theta_H)$, $\eta=(\eta_1, \ldots, \eta_n)^{\T}$  and $y_j=(y_{1j}, \ldots, y_{nj})^{\T}$;
	}
	\nl \For{j from 1 to p}{
		sample $\sigma_j^2$ from $\mbox{InvGa}\{a_\sigma+0.5 n,  b_\sigma+0.5 \sum_{i=1}^{n} (y_{ij}-\sum_{h=1}^{H}\lambda_{jh} \eta_{ih})^2\}$;
	}
	\nl \For{i from 1 to n}{
		sample $\eta_i$ from $N_H \{(I_H+\Lambda^{\T} \Sigma^{-1} \Lambda)^{-1}\Lambda^{\T} \Sigma^{-1} y_i, (I_H+\Lambda^{\T} \Sigma^{-1} \Lambda)^{-1} \}$;
	}
	\nl \For{h from 1 to H}{
		sample $z_h$ from the categorical distribution with probabilities as in~\eqref{eq_discr_1}; 
	}
	\nl \For{l from 1 to $ (H-1) $}{
		update $v_l$ from $\mbox{Beta}\{1+\sum_{h=1}^H \mathds{1}(z_h=l), \alpha+\sum_{h=1}^H \mathds{1}(z_h>l)\}$;
	}
	set $v_H=1$ and update $\omega_1, \ldots, \omega_H$ from $v_1, \ldots, v_H$ through~\eqref{eq_cusp_fact};\\
	\nl \For{h from 1 to H}{
		\textbf{if} $z_h \leq h$ \textbf{then}  $\theta_h=\theta_{\infty}$ \textbf{else} sample $\theta_h$ from $\mbox{InvGa}(a_{\theta}+0.5 p,b_{\theta}+0.5 \sum_{j=1}^p \lambda^2_{jh})$;
	}
	\textbf{Output at the end of one cycle:} one sample from the posterior of $ \Omega = \Lambda \Lambda^\T + \Sigma $.
	\vspace{-5pt}
\end{algorithm}

\begin{algorithm}[t]
	\caption{One cycle of the adaptive  version for the Gibbs sampler in Algorithm~\ref{alg_factor}} 
	\label{alg_adaptive}
	\vspace{-7pt}
	Let $t$ be the cycle number, $H^{(t)}$ the truncation index at $t$, and $ H^{*(t)}=\sum_{h=1}^{H^{(t)}} \mathds{1}(z^{(t)}_h>h) $. \\
	\nl Perform one cycle of Algorithm~\ref{alg_factor};\\
	\nl  \If{$ t \geq \bar{t} $}{adapt with probability $ p(t)=\exp(\alpha_0+\alpha_1 t)$ as follows\\
		\nl  \eIf{$ H^{*(t)} < H^{(t-1)}-1 $}{
			set $ H^{(t)}=H^{*(t)}+1 $, drop the inactive columns in $ \Lambda$ together with the associated parameters in $\eta, \theta, w$, and add a final component to $ \Lambda, \eta, \theta, w$ sampled from the prior;}{
			set $ H^{(t)}=H^{(t-1)}{+}1 $ and add a final column sampled from the spike to $ \Lambda$, together with the associated parameters in $\eta, \theta $ and $w$, sampled from the corresponding priors;}}			
	\textbf{Output:} one sample from the posterior of $ \Omega = \Lambda \Lambda^\T + \Sigma $ and a value for $H^*$.
	\vspace{-5pt}
\end{algorithm}

\subsection{Tuning the truncation index via adaptive Gibbs sampling}
\label{subsec_factor_adaptive}

Recalling \S~\ref{subsec_factor_formulation}, it is reasonable to perform Bayesian inference with at most $p$ factors. Under our cumulative shrinkage process truncated at $ H $ terms this translates into $ H \leq p+1 $, since there are at most $ H-1 $ active factors, with the $H$th one modelled with the spike by construction. However, this choice  is too conservative, since we expect substantially fewer active factors than $ p $, especially when $ p $ is very large. Hence, running Algorithm~\ref{alg_factor} with $ H = p+1 $ would be computationally inefficient, since most of the columns in   $\Lambda$  would be modelled by the spike, thus providing a negligible contribution to the factorization of $\Omega$.

\citet{bhattacharya2011sparse} addressed this issue via an adaptive Gibbs sampler which tunes~$ H $ as the sampler proceeds.  To satisfy the diminishing adaptation condition in \citet{roberts2007coupling}, they adapt $ H $ at the iteration $ t $ with probability $ p(t)=\exp(\alpha_0+\alpha_1 t) $, where $ \alpha_0 \leq 0 $ and $ \alpha_1 < 0 $.  This adaptation consists in dropping the inactive columns of $\Lambda$, if any, together with the corresponding parameters. If instead all columns are active, an extra factor is added, sampling the associated parameters from the prior.

This idea can be also implemented for the cumulative shrinkage process, as illustrated in Algorithm~\ref{alg_adaptive}. Under our prior, the inactive $\Lambda$ columns are naturally identified as those modelled by the spike and, hence, have index $ h $ such that $ z_h \leq h $. Under the multiplicative gamma process, instead, a column is flagged as inactive if all its entries are within distance $ \epsilon $ from zero.  This $ \epsilon $ plays a similar role as our spike location $ \theta_\infty $. Indeed, lower values of $ \epsilon $ and $ \theta_\infty $ make it harder to discard inactive columns, thus affecting running time. Hence, although fixing $ \theta_\infty $ close to zero is a key to enforce shrinkage, excessively low values should be avoided. Since under a truncated cumulative shrinkage process  the number of active factors $ H^* $ is at most $ H-1 $, we increase $ H $ by one when $ H^* = H -1 $, and we decrease $ H $ to $ H^* + 1 $ when $ H^* < H-1 $.

In our implementation no adaptation is allowed before a fixed number $\bar{t}$ of iterations to let the chain stabilize, while $H$ and $H^*$ are initialized to $p+1$ and $p$, which is the maximum possible rank for $\Omega$. Further guidance for the choice of $ H $ can be obtained by monitoring how close $ E(\pi_H) $ is to 1, via \eqref{eq_cusp_mean}.

\begin{table}[b]
	\def~{\hphantom{0}}
	\caption{Performance of \textsc{cusp} and \textsc{mgp}  in $ 25 $ simulations for different  $ (p,H_0) $ scenarios}{
		\begin{tabular*}{0.8\textwidth}{rlcccccc}
			$ (p,H_0) $ & method & \multicolumn{2}{c}{\textsc{mse}} & \multicolumn{2}{c}{$ E(H^* \mid y) $} & averaged \textsc{ess} & runtime (s) \\
			& & median & \textsc{iqr} & median & \textsc{iqr} & median & median \\ 
			(20,5) & \textsc{cusp} & 0.75 & 0.29 & 5.00 & 0.00 & 655.04 & 310.76  \\ 
			& \textsc{mgp} & 0.75 & 0.32 & 19.69 & 0.21 & 547.23 & 616.61 \\ 
			(50,10) & \textsc{cusp} & 2.25 & 0.33 & 10.00 & 0.00 & 273.55 & 716.23  \\ 
			& \textsc{mgp} & 2.26 & 0.28 & 28.64 & 1.94 & 251.35 & 1845.88  \\ 
			(100,15) & \textsc{cusp} & 3.76 & 0.40 & 15.00 & 0.00 & 175.26 & 2284.87  \\ 
			& \textsc{mgp} & 3.97 & 0.45 & 34.38 & 2.92 & 116.10 & 5002.33  \\ 
	\end{tabular*}}
	\label{table_compare}
	\caption*{
		\textsc{cusp},~cumulative shrinkage process; 
		\textsc{mgp},~multiplicative gamma process;
		\textsc{mse},~mean square error;
		\textsc{ess},~effective sample size;
		\textsc{iqr},~interquartile range.
	}
\end{table}

\section{Performance assessments of Gaussian factor models in simulations}
\label{subsec_factor_empirical}
We consider illustrative simulations to assess performance  in learning the structure of the true covariance matrix $\Omega_0=\Lambda_0 \Lambda^{\T}_0+ \Sigma_0$ for the data $y=(y_1, \ldots, y_n)$ from a Gaussian factor model, with $ \Sigma_0 = I_{p} $ and the entries in $ \Lambda_0  \in \Re^{p \times H_0}$ drawn from independent $ N(0,1) $.  To study performance at varying dimensions, we consider three different combinations  of $(p,H_0)$: $(20,5) $, $ (50,10) $ and $ (100,15) $.    For every  pair $ (p,H_0) $, we sample 25 datasets of $n=100$ observations from $N_{p}(0,\Omega_0)$ and, for each of the 25 replicates, we perform posterior inference on $\Omega$ via the Gaussian factor model in \S~\ref{subsec_factor_formulation} under both prior~\eqref{eq_mgp} and~\eqref{eq_cusp_fact}, exploiting the adaptive Gibbs sampler in \cite{bhattacharya2011sparse} and Algorithm~\ref{alg_adaptive}, respectively.

\begin{table}[t]
	\def~{\hphantom{0}}
	\caption{Sensitivity analysis for \textsc{cusp} hyper-parameters $ (\alpha,a_\theta,b_\theta,\theta_\infty) $ in $ 25 $ simulations}{
		\begin{tabular*}{1\textwidth}{rlcccccc}
			$ (p,H_0) $ & $ (\alpha,a_\theta,b_\theta,\theta_\infty) $ & \multicolumn{2}{c}{\textsc{mse}} & \multicolumn{2}{c}{$ E(H^* \mid y) $} & averaged \textsc{ess} & runtime (s) \\
			& & median & \textsc{iqr} & median & \textsc{iqr} & median & median\\ 
			(20,5) & (2.5,2,2,0.05) & 0.74 & 0.32 & 5.00 & 0.00  & 626.22 & 317.31 \\ 
			& (10,2,2,0.05) & 0.74 & 0.33 & 5.00 & 0.00  & 636.61 & 314.82 \\ 
			& (5,2,1,0.05) & 0.72 & 0.34 & 5.00 & 0.00  & 607.61 & 322.68 \\ 
			& (5,1,2,0.05) & 0.79 & 0.30 & 5.00 & 0.00  & 602.28 & 309.39 \\ 
			& (5,2,2,0.025) & 0.78 & 0.31 & 5.00 & 0.00 & 655.80 & 313.21 \\ 
			& (5,2,2,0.1) & 0.74 & 0.30 & 5.00 & 0.04  & 604.88 & 315.51 \\ 
			(50,10) & (2.5,2,2,0.05) & 2.25 & 0.40 & 10.00 & 0.00 & 280.39 & 719.11 \\ 
			& (10,2,2,0.05) & 2.20 & 0.36 & 10.00 & 0.00 & 277.89 & 748.75 \\ 
			& (5,2,1,0.05) & 2.16 & 0.42 & 10.00 & 0.00  & 266.82 & 722.67 \\ 
			& (5,1,2,0.05) & 2.35 & 0.40 & 10.00 & 0.00  & 272.47 & 689.70 \\ 
			& (5,2,2,0.025) & 2.22 & 0.35 & 10.00 & 0.00 & 280.60 & 717.19 \\ 
			& (5,2,2,0.1) & 2.22 & 0.41 & 10.00 & 0.00 & 273.39 & 698.96 \\ 
			(100,15) & (2.5,2,2,0.05) & 3.68 & 0.47 & 15.00 & 0.00 & 176.31 & 2247.44 \\ 
			& (10,2,2,0.05) & 3.74 & 0.40 & 15.00 & 0.00  & 172.02 & 2205.78 \\ 
			& (5,2,1,0.05) & 3.64 & 0.44 & 15.00 & 0.00 & 172.04 & 2287.32 \\ 
			& (5,1,2,0.05) & 3.96 & 0.52 & 15.00 & 0.00  & 174.74 & 2178.47 \\ 
			& (5,2,2,0.025) & 3.70 & 0.44 & 15.00 & 0.00 & 172.83 & 2200.20 \\ 
			& (5,2,2,0.1) & 3.77 & 0.44 & 15.00 & 0.00 & 174.76 & 2284.80 \\ 
	\end{tabular*}}
	\label{table_sens}
	\caption*{
		\textsc{cusp},~cumulative shrinkage process; 
		\textsc{mse},~mean square error;
		\textsc{ess},~effective sample size;
		\textsc{iqr},~interquartile range.
	}
\end{table}

For our cumulative shrinkage process, we set $\alpha=5$, $a_\theta=b_\theta=2$ and $\theta_\infty=0.05$, whereas for the multiplicative gamma process, we follow \citet{durante2017note} by considering $(a_1, a_2)=(1,2)$, and set $\nu=3$ as done by  \citet{bhattacharya2011sparse} in their simulations.  For both models, $(a_\sigma, b_\sigma)$ are fixed at $(1, 0.3)$ as in \cite{bhattacharya2011sparse}. The truncation $ H $ is initialized at $ p $ for the multiplicative gamma process and at $ p+1 $ for the cumulative shrinkage process, both corresponding to at most $ p $ active factors. For the two methods, adaptation is allowed only after $ 500 $ iterations and, following \citet{bhattacharya2011sparse}, the parameters $ (\alpha_0,\alpha_1) $ are set to $ (-1,-5 \times 10^{-4}) $, while the adaptation threshold $ \epsilon $ in the multiplicative gamma process is $ 10^{-4} $. Both algorithms are run for $10000$ iterations  after a burn-in of $5000$ and, by thinning every 5, we obtain a final sample of $ 2000 $ draws from the posterior of $\Omega$.  For each of the $ 25 $ simulations in every scenario, we compute a Monte Carlo estimate of  $\sum_{j=1}^p\sum_{q=j}^pE\{(\Omega_{jq}-\Omega_{0jq})^2\mid y\}/ \{p(p+1)/2 \}$ and $ E(H^* \mid y) $. Since $E\{(\Omega_{jq}-\Omega_{0jq})^2\mid y\}=\{E(\Omega_{jq} \mid y)-\Omega_{0jq}\}^2+\mbox{var}(\Omega_{jq} \mid y)$, the posterior averaged mean square error accounts for both bias and variance in the posterior of $\Omega$.

Table~\ref{table_compare} shows, for each scenario and model, the median and the interquartile range of the above quantities computed from the $25$ measures produced by the different simulations, together with the medians of the averaged effective sample sizes, out of $2000$ samples, and of the running times.  Such quantities rely on an \texttt{R} implementation run on an Intel~Core i7-3632QM~CPU laptop with $ 7.7 $~GB of RAM. The two methods have comparable mean square errors, but these measures and the performance gains of prior  \eqref{eq_cusp_fact} over  \eqref{eq_mgp} increase with $H_0$.  Our  approach also provides some improvements in mixing and reduced running times. The latter is arguably due to the fact that the multiplicative gamma process  overestimates $H^*$, hence keeping more parameters to update  than necessary. Instead, our cumulative shrinkage process recovers the true dimension $ H_0 $ in all settings, thus efficiently tuning the truncation level $H$.  Such an improved learning of the true underlying dimension is confirmed by the $95\%$ credible intervals highly concentrated around $H_0$ in all the scenarios considered. The multiplicative gamma process leads instead to wider credible intervals for $H^*$, with none of them including $H_0$.  As shown in Table~\ref{table_sens}, results are robust to moderate and reasonable changes in the hyper-parameters of the cumulative shrinkage process. We also tried to modify $ \epsilon $ in \citet{bhattacharya2011sparse} so as to delete $\Lambda$ columns with values on the same scale of our spike. This setting provided lower estimates for $H^*$ and, hence, a computational time more similar to our cumulative shrinkage process, but led to worse mean square errors and still some difficulties in learning $H_0$. 

\section{Application of  Gaussian factor models to personality data}
\label{sec_app}
We conclude with an application to a subset of the personality data available in the dataset \texttt{bfi} from the \texttt{R} package \texttt{psych}. Here, we focus on  the association structure among $p=25$ personality self-report items collected on a 6 point response scale for $n=126$ individuals older than $50$ years. These variables represent answers to questions organized into five  personality traits known as agreeableness, conscientiousness, extraversion, neuroticism, and openness. Recalling common implementations of factor models, we center the 25 items, and then replace variables $1,9,10,11,12,22$ and $25$ with their negative version as suggested in the  \texttt{R} documentation of the \texttt{bfi}  dataset  to have  coherent answers within each personality trait. Posterior inference under priors  \eqref{eq_mgp}--\eqref{eq_cusp_fact} is performed with the same hyper-parameters and Gibbs settings as in \S~\ref{subsec_factor_empirical}.

\begin{figure}[t]
	\captionsetup{font={small}}
	\begin{center}
		\includegraphics[height=4.1cm]{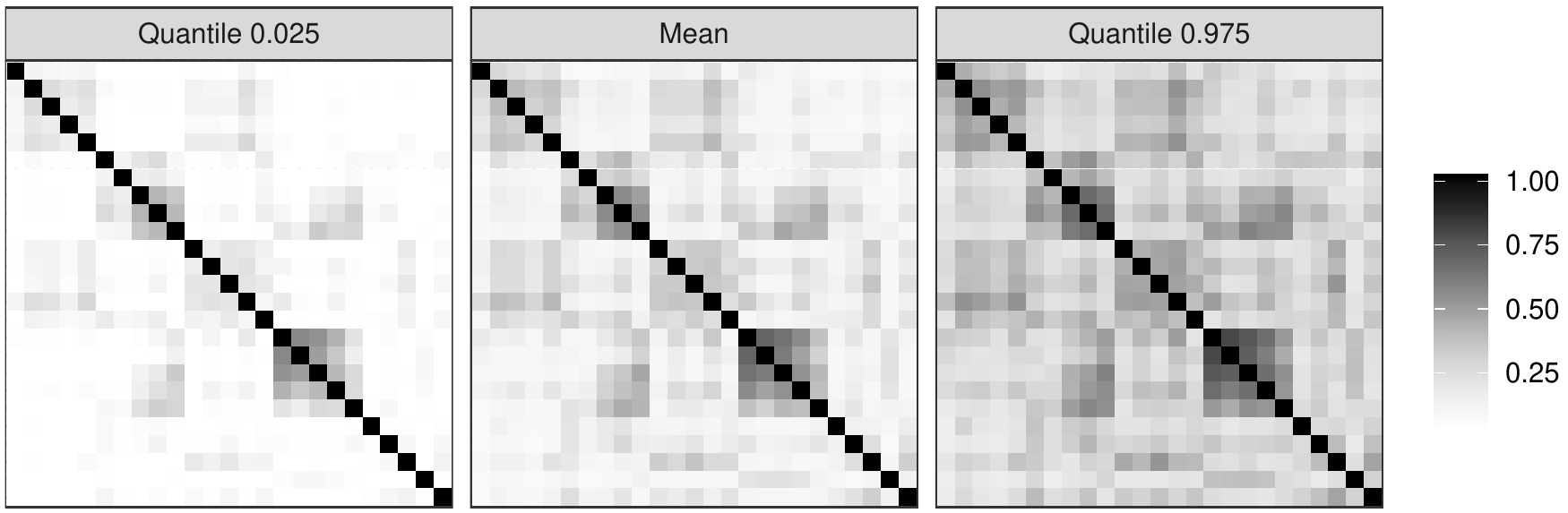}
	\end{center}
	\vspace{-6pt}
	\caption{Posterior mean and credible intervals for each element  of the absolute correlation matrix $|\bar{\Omega}|$ under our model.}
	\label{f1}
\end{figure}

Figure \ref{f1} shows posterior means and credible intervals for the absolute value of the entries in the correlation matrix $\bar{\Omega}$, under our model. Samples from $\bar{\Omega}$ are obtained computing $\bar{\Omega}=(\Omega \odot I_p)^{-\frac{1}{2}} \Omega (\Omega \odot I_p)^{-\frac{1}{2}}$ for every sample of  $\Omega=\Lambda \Lambda^{\T}+\Sigma$, with $\odot$ denoting the element-wise Hadamard product. Figure \ref{f1} highlights associations within each block of five answers measuring a main personality trait, while showing also  interesting across-blocks correlations among agreeableness and extraversion as well as conscientiousness and neuroticism. Openness has less evident within-block and across-block associations. These results suggest three main factors as confirmed by the posterior mean and by the $95\%$ credible intervals for $H^*$ under the cumulative shrinkage process, which are $2.84$ and $(2,3)$, respectively. Such posterior summaries are  $24.01$ and $(18,25)$ under the multiplicative gamma process, but the higher $H^*$ does not lead to improved learning of $\bar{\Omega}$. In fact, when considering the Monte Carlo estimate of the mean squared deviations $\sum_{j=1}^p\sum_{q=j}^pE(\bar{\Omega}_{jq}-S_{jq})^2/ \{p(p+1)/2 \}$  from the sample correlation matrix $S$, we obtain $0.01$ under both \eqref{eq_cusp_fact} and \eqref{eq_mgp}, suggesting that the multiplicative gamma process might overestimate $H^*$ in this application. This leads to more redundant parameters to be updated in the adaptive Gibbs sampler, thus increasing the computational time from $400.69$  to $1321.04$ seconds. Our approach also increases the averaged effective sample size from $901.68$ to $1070.83$.


\section*{Acknowledgement}
The authors are grateful to the Editor, the Associate Editor and the referees for the useful suggestions, and acknowledge the support from \textsc{miur} (\textsc{prin} 2017 grant) as well as the United States Office of Naval Research and National Institutes of Health in the preparation of the final version of this article.

\appendix
\section*{Appendix}
\begin{proof}[of Proposition~\ref{prop_step}.]
	Since the mapping from the sequence $w=\{w_h \in (0,1): h=1, 2, \ldots \}$ to $\pi=\{\pi_h \in (0,1): h=1, 2, \ldots \}$ is one-to-one, it is sufficient to ensure that the stick-breaking prior for $w$ has full support on the infinite dimensional simplex. This result is proved by \citet{bissiri_2014} in \S~3.2.
\end{proof}

\begin{proof}[of Proposition~\ref{lemma_dtv}.] The proof of Proposition~2 adapts the one of Theorem~1 in \cite{canale2017convex}. In fact, under the prior in Definition~1, the distance $d_{\textsc{tv}}(P_0,P_h)$ on the Borel $ \sigma $-algebra in $ \Re $ is equal to
	\begin{eqnarray*}
		\sup_{\mathbb{A} \in \mathcal{B}(\Re)} \vert P_0(\mathbb{A}) - P_h(\mathbb{A}) \vert {=} \sup_{\mathbb{A} \in \mathcal{B}(\Re)} \vert P_0(\mathbb{A}) - ( 1 - \pi_h)P_0(\mathbb{A}) - \pi_h \delta_{\theta_{\infty}}(\mathbb{A}) \vert{=} \ \pi_h\sup_{\mathbb{A} \in \mathcal{B}(\Re)} \vert P_0(\mathbb{A}) - \delta_{\theta_{\infty}}(\mathbb{A}) \vert.
	\end{eqnarray*}
	Hence $d_{\textsc{tv}}(P_0,P_h)=\pi_h d_{\textsc{tv}}(P_0,\delta_{\theta_{\infty}})$, completing the proof.
\end{proof}

\begin{proof}[of Lemma~\ref{lemma_ordering}.] Notice that, for each $h$, $\mbox{\normalfont pr}(|\theta_h-\theta_{\infty}| > \varepsilon)$ can be equivalently expressed as
	$$E[P_h\{\bar{\mathbb{B}}_{\varepsilon}(\theta_\infty)\}] = 
	E[( 1 - \pi_h ) P_0\{\bar{\mathbb{B}}_{\varepsilon}(\theta_\infty)\} + \pi_h  \delta_{\theta_{\infty}}\{\bar{\mathbb{B}}_{\varepsilon}(\theta_\infty)\}] =
	P_0\{\bar{\mathbb{B}}_{\varepsilon}(\theta_\infty)\}\{1- E(\pi_h)\}.$$
	Therefore, replacing $E(\pi_h)$ with its expression in  equation~(2) leads to~(4). To prove that $\mbox{\normalfont pr}(|\theta_{h+1}{-}  \theta_{\infty}| \leq \varepsilon) > \mbox{\normalfont pr}(|\theta_h{-}  \theta_{\infty}| \leq \varepsilon)$ it is sufficient to note that $\{\alpha(1+\alpha)^{-1}\}^{h+1}<\{\alpha(1+\alpha)^{-1}\}^h$.
\end{proof}

\begin{proof}[of Theorem~\ref{thm_trunc2}.] The proof follows after noting that
	$ \mbox{pr}(\sup_{h>H} |\theta_h| > \varepsilon ) = \mbox{\normalfont pr}\{\cup_{h>H}(|\theta_h|>\varepsilon)\}$, and that $\delta_{\theta_{\infty}}\{\bar{\mathbb{B}}_{\varepsilon}(0)\}=0$ for any $\varepsilon \geq |\theta_{\infty}|$.  Hence, adapting the proof of Lemma~1, we obtain
	\begin{eqnarray*}
		\mbox{\normalfont pr}\{\cup_{h>H}(|\theta_h|>\varepsilon)\}\leq \sum\nolimits_{h=H+1}^{\infty}\mbox{\normalfont pr}(|\theta_h| > \varepsilon)
		=P_0\{\bar{\mathbb{B}}_{\varepsilon}(0)\} \sum\nolimits_{h=H+1}^{\infty}\{\alpha(1+\alpha)^{-1}\}^h.
	\end{eqnarray*}
	To conclude the proof, notice that 
	$\sum_{h=H+1}^{\infty}\{\alpha(1+\alpha)^{-1}\}^h = \alpha\{\alpha(1+\alpha)^{-1}\}^H$. 
\end{proof}


\begin{proof}[of Theorem~\ref{thm_support}.] 
	Let us first prove that for the Gaussian factor model in \S~3{$\cdot$}1, with prior (6) truncated at $H$ terms, we have  $ {\Pi \{ \Omega \in \Re^{p \times p}: \Omega\mbox{ has finite entries and is positive semi-definite} \} = 1 }$. Since, by construction, $ \Sigma $ is diagonal with almost surely finite and non-negative entries, and  $\Lambda \Lambda^{\T} $ is trivially positive semi-definite, we only need to ensure that each entry $ \lambda_{r\cdot} \lambda_{j\cdot}^{\T} $ in $\Lambda \Lambda^{\T} $ is almost surely finite. By the Cauchy-Schwartz inequality we obtain $$ | \lambda_{r\cdot} \lambda_{j\cdot}^{\T}| \leq \| \lambda_{r\cdot} \| \| \lambda_{j\cdot} \| \leq \max_{1 \leq j \leq p} \| \lambda_{j\cdot} \|^2.$$
	Under the factor model in \S~$3{\cdot}1$ having prior (6) truncated at $H$ terms, we have that $$ E(\| \lambda_{j\cdot} \|^2) = \sum\nolimits_{h=1}^{H} E(\lambda_{jh}^2) = 
	\sum\nolimits_{h=1}^{H} E\{ E(\lambda_{jh}^2 \mid \theta_h) \} =
	\sum\nolimits_{h=1}^{H} E( \theta_h ), $$ 
	for every $j=1, \ldots, p$, including the index of the maximum, thus ensuring that each entry in $\Lambda \Lambda^{\T} $ is almost surely finite under the sufficient condition that $E( \theta_h )< \infty$ $(h=1, \ldots, H)$. This holds when $a_{\theta}>1$ and $\theta_{\infty}<\infty$.

	Let us now prove the full support for $\Pi$. Since $H>H_0$, there always exists a $\Lambda \in \Re^{p \times H}$ and a positive diagonal matrix $\Sigma$ such that $\Lambda \Lambda^\T + \Sigma = \Lambda_0 \Lambda_0^\T + \Sigma_0$. For instance, one can let $\Sigma=\Sigma_0$ and $\Lambda=[\Lambda_0,0_{p \times (H-H_0)}]$. Hence, it suffices to prove full support for the priors induced on $\Lambda$ and $\Sigma$ by the truncated version of our cumulative shrinkage process. Such a property easily holds for $\Sigma$, whose diagonal elements $\sigma_j^2 \ (j=1,\ldots,p)$ have independent inverse-gamma priors. Moreover, adapting the proof of Proposition 2 in  \citet{bhattacharya2011sparse}, full support can be proved also for the prior induced on $\Lambda$. Indeed, recalling \S~$3{\cdot}1$, we have that $ \mbox{pr} \{ \sum_{j=1}^{p} \sum_{h=1}^{H} (\lambda_{jh}-\lambda_{0jh})^2 < \epsilon_1^2 \} \geq \mbox{pr} \{ (\lambda_{jh}-\lambda_{0jh})^2 < \epsilon_1^2 / (pH), \mbox{ for all } j=1,\ldots,p; \ h=1,\ldots,H \}$ with
	\begin{eqnarray*}
		&&\mbox{pr} \{ (\lambda_{jh}-\lambda_{0jh})^2 < \epsilon_1^2 / (pH), \mbox{ for all } j=1,\ldots,p; \ h=1,\ldots,H \} \\
		&&= E \left[ \prod_{j=1}^{p} \prod_{h=1}^{H} \mbox{pr} \{ (\lambda_{jh}-\lambda_{0jh})^2 < \epsilon_1^2 / (pH) | \theta \} \right] >0.
	\end{eqnarray*}
	In fact, conditioned on $ \theta =(\theta_1,\ldots,\theta_H) $, each $ \lambda_{jh} $ has independent $ N(0,\theta_h) $ distribution.
\end{proof}

\bibliographystyle{apalike}
\bibliography{shrinkage}

\begin{thebibliography}{}

\bibitem[Bhattacharya and Dunson, 2011]{bhattacharya2011sparse}
Bhattacharya, A. and Dunson, D.~B. (2011).
\newblock Sparse {Bayesian} infinite factor models.
\newblock {\em Biometrika}, 98:291--306.

\bibitem[Bickel and Levina, 2008]{bickel_2008}
Bickel, P.~J. and Levina, E. (2008).
\newblock Regularized estimation of large covariance matrices.
\newblock {\em Ann. Statist.}, 36(1):199--227.

\bibitem[Bissiri and Ongaro, 2014]{bissiri_2014}
Bissiri, P.~G. and Ongaro, A. (2014).
\newblock On the topological support of species sampling priors.
\newblock {\em Electron. J. Statist.}, 8(1):861--882.

\bibitem[Canale et~al., 2018]{canale2017convex}
Canale, A., Durante, D., and Dunson, D.~B. (2018).
\newblock Convex mixture regression for quantitative risk assessment.
\newblock {\em Biometrics}, 74:1331--1340.

\bibitem[Carvalho et~al., 2008]{carvalho_2008}
Carvalho, C.~M., Chang, J., Lucas, J.~E., Nevins, J.~R., Wang, Q., and West, M.
  (2008).
\newblock High-dimensional sparse factor modeling: applications in gene
  expression genomics.
\newblock {\em J. Am. Statist. Assoc.}, 103:1438--1456.

\bibitem[Carvalho et~al., 2010]{carvalho2010}
Carvalho, C.~M., Polson, N.~G., and Scott, J.~G. (2010).
\newblock The horseshoe estimator for sparse signals.
\newblock {\em Biometrika}, 97:465--480.

\bibitem[Durante, 2017]{durante2017note}
Durante, D. (2017).
\newblock A note on the multiplicative gamma process.
\newblock {\em Statist. Probabil. Lett.}, 122:198--204.

\bibitem[Gopalan et~al., 2014]{gopalan2014bayesian}
Gopalan, P., Ruiz, F.~J., Ranganath, R., and Blei, D. (2014).
\newblock Bayesian nonparametric {P}oisson factorization for recommendation
  systems.
\newblock {\em J. Mach. Learn. Res. W\&CP}, 33:275--283.

\bibitem[Ishwaran and James, 2001]{ishwaran_2001}
Ishwaran, H. and James, L.~F. (2001).
\newblock Gibbs sampling methods for stick-breaking priors.
\newblock {\em J. Am. Statist. Assoc.}, 96:161--173.

\bibitem[Ishwaran and Rao, 2005]{ishwaran2005spike}
Ishwaran, H. and Rao, J.~S. (2005).
\newblock Spike and slab variable selection: frequentist and {Bayesian}
  strategies.
\newblock {\em Ann. Statist.}, 33:730--773.

\bibitem[Knowles and Ghahramani, 2011]{knowles_2011}
Knowles, D. and Ghahramani, Z. (2011).
\newblock Nonparametric {Bayesian} sparse factor models with application to
  gene expression modeling.
\newblock {\em Ann. Appl. Statist.}, 5:1534--1552.

\bibitem[Lopes and West, 2004]{lopes2004}
Lopes, H.~F. and West, M. (2004).
\newblock Bayesian model assessment in factor analysis.
\newblock {\em Statist. Sinica}, 14(1):41--68.

\bibitem[Roberts and Rosenthal, 2007]{roberts2007coupling}
Roberts, G.~O. and Rosenthal, J.~S. (2007).
\newblock Coupling and ergodicity of adaptive {Markov} chain {Monte} {Carlo}
  algorithms.
\newblock {\em J. Appl. Prob.}, 44(2):458--475.

\bibitem[Rousseau and Mengersen, 2011]{rousseau2011}
Rousseau, J. and Mengersen, K. (2011).
\newblock Asymptotic behaviour of the posterior distribution in overfitted
  mixture models.
\newblock {\em J. R. Statist. Soc. {\normalfont B}}, 73:689--710.

\bibitem[Schwartz, 1965]{schwartz1965bayes}
Schwartz, L. (1965).
\newblock On {Bayes} procedures.
\newblock {\em Prob. Theory Rel. Fields}, 4(1):10--26.

\bibitem[Tibshirani, 1996]{tibshirani1996}
Tibshirani, R. (1996).
\newblock Regression shrinkage and selection via the lasso.
\newblock {\em J. R. Statist. Soc. {\normalfont B}}, 58(1):267--288.

\bibitem[Zou and Hastie, 2005]{zou2005}
Zou, H. and Hastie, T. (2005).
\newblock Regularization and variable selection via the elastic net.
\newblock {\em J. R. Statist. Soc. {\normalfont B}}, 67(2):301--320.

\end{thebibliography}

\end{document}